\documentstyle[11pt,paspconf,epsf]{article}

\begin{document}

\title { Probing Star Formation Timescales in Elliptical Galaxies}
\author{Daniel Thomas}
\affil{Universit\"ats-Sternwarte, Scheinerstr.\ 1, D-81679 M\"unchen, Germany}
\author{Guinevere Kauffmann}
\affil{Max-Planck-Institut f\"ur Astro-Physik, D-85740 Garching, Germany}


\begin{abstract}

In models of galaxy formation in a hierarchical Universe, elliptical
galaxies form through the merging of smaller disk systems. These models
yield a number of testable predictions if reliable techniques for
determining the relative ages and compositions of the stellar populations
of different galaxies can be found: 1) ellipticals in low-density
environments form later than ellipticals in clusters, 2) more massive
ellipticals form later, 3) more massive ellipticals form in
dissipationless mergers from disk galaxies with low gas content. While
colours and the Balmer line strengths of galaxies can be used to infer
the average ages of the stellar populations of ellipticals, $\alpha$/Fe
element ratios carry information about the timescale over which star
formation took place. Here we present preliminary results from
semi-analytic models for the distribution of Mg/Fe ratios in galaxies as
a function of morphological type, luminosity and environment.

\end{abstract}



\section {Hierarchical Clustering and the Formation of Ellipticals 
through Mergers}

According to the standard theoretical paradigm, the structures observed
in the Universe today were formed by the gravitational amplification of
small perturbations in an initially Gaussian dark matter density field.
Small scale overdensities were the first to collapse, and the resulting
objects subsequently merged under the influence of gravity to form larger
structures such as groups and clusters of galaxies. Galaxies formed
within dense {\em halos} of dark matter, where gas was able to reach high
enough overdensities to cool, condense and form stars.

The quiescent cooling of gas within a dark matter halo results in the
formation of a rotationally-supported disk system at the centre of the
halo. When halos merge with each other, a bound group of galaxies is
produced. Dynamical friction will cause the orbits of the group members
to erode over time, and the galaxies to spiral in towards the centre of
the halo and merge. Galaxy-galaxy mergers are thus inevitable
in this picture. N-body simulations have demonstrated that mergers
between disk galaxies of near-equal mass result in the formation of
remnant systems that are structurally very similar to observed elliptical
galaxies (e.g. Barnes \& Hernquist 1996).

Elliptical galaxies are a very homogeneous class of objects. Their
stellar populations are old and in addition there is a well-defined
correlation between  colour and absolute magnitude that exhibits a
remarkably small scatter (Bower, Lucey \& Ellis 1992). There has been
some doubt as to whether the old ages of ellipticals and the small
scatter of the colour-magnitude relation can be explained in a scenario
in which all ellipticals form by mergers of spirals. In order to address
this issue, one must follow the formation and evolution of elliptical
galaxies in the framework of the theoretical scenario described above.

\section {Semi-Analytic Models of Galaxy Formation}   

In semi-analytic models of galaxy formation (see for example Kauffmann,
White \& Guiderdoni 1993; Cole et al 1994; Somerville \& Primack 1999),
an algorithm based on the extended Press-Schechter theory is used to
generate Monte Carlo realizations of the merging paths of dark matter
halos from high redshift to the present. These ``merger trees'' are
specified for a chosen set of cosmological initial conditions and allow
the progenitors of a present-day dark matter halo to be traced back to
arbitrarily early times. Simple recipes are introduced to describe the
cooling of gas within the halos, the formation of stars from the cold
gas, feedback and heavy element enrichment from supernova explosions, and
the merging rates of galaxies.
The models are then coupled to a stellar population synthesis code in
order to generate magnitudes and colours that can be compared directly
with observational data.

If two galaxies of comparable mass merge, the stars of both objects are
added together to create a bulge component. When a bulge is formed by a
merger, the cold gas present in the two galaxies is transformed into
stars in a ``starburst'' with a timescale of $10^8$ years. Further
cooling of gas in the halo may lead to the formation of a new disk. The
morphological classification of galaxies is made according their B-band
disk-to-bulge ratios (Simien \& de Vaucouleurs 1986). If
$M(B)_{bulge}-M(B)_{total} < 1$ mag, then the galaxy is classified as
early-type (elliptical or S0).

\section { When, Where and How do Ellipticals Form in the Merger Picture?}

\begin{figure}[ht!]
\begin{center}
\begin{minipage}{7cm}
\plotone{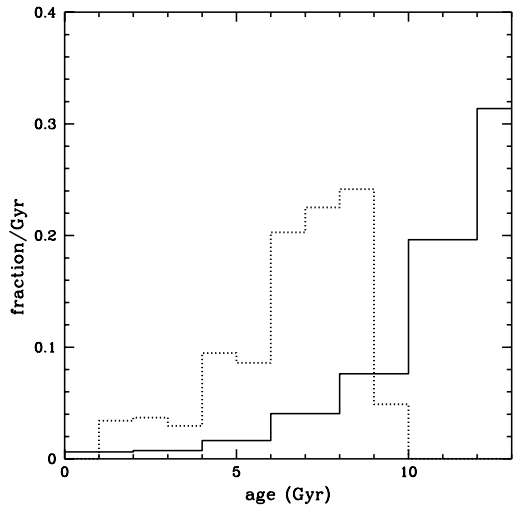}
\end{minipage}
\end{center}
\caption{The solid line
shows the {\em average} star formation history of a present-day cluster
elliptical galaxy. The dotted line shows the distribution of the time of
the last major merger. Results are shown for a high-density cold dark
matter (CDM) cosmology with $\Omega=1$, $H_0 = 50$ km s$^{-1}$ Mpc$^{-1}$
and $\sigma_8 =0.67$.}
\label{fig:sfh}
\end{figure}

It is important to define carefully what is meant by the ``formation
time'' of an elliptical galaxy.  The solid line in Fig.~\ref{fig:sfh}
shows the {\em average} star formation history of a present-day cluster
elliptical galaxy. Results are shown for a high-density cold dark matter
(CDM) cosmology with $\Omega=1$, $H_0 = 50$ km s$^{-1}$ Mpc$^{-1}$ and
$\sigma_8 =0.67$. As can be seen, most of the stars form at redshifts
greater than 2. Because the stars form early, the small scatter of the
observed colour-magnitude relation is not a problem for the model
(Kauffmann 1996). The dotted line shows the distribution of the time of
the last major merger. This is typically much later than the epoch at
which the stars form; in this cosmology, many elliptical galaxies have
their last major mergers at redshifts less than 1. The epoch of the last
major merger also depends on the choice of cosmological parameters. In a
low density model ($\Omega < 1$), the epoch at which structure forms is
shifted to higher redshift. At z=1, more than $70 \%$ of present-day
bright ellipticals are already present in a CDM model with $\Omega=0.3$
and $\Lambda=0.7$ (Kauffmann et al 1999), but most have not yet formed by
$z\sim 2$. Finally, the formation time of an elliptical depends on its
mass. More massive objects form later. Because the ratio of gas
to stellar mass in galaxies decreases at late times, massive ellipticals
form from gas-poor progenitors and form a smaller fraction of their stars
in the merger-induced starburst (Kauffmann \& Haehnelt, in preparation).

The probability of merging peaks in environments where the relative
velocities between galaxies are comparable to their internal stellar
velocity dispersions. In the semi-analytic models, bright ($ > L_*$)
elliptical galaxies are formed in dark matter halos with circular
velocities in the range 250--350 km s$^{-1}$. These newly-formed
ellipticals are {\em isolated} galaxies, with no bright companion within
a radius of $\sim 1$ Mpc. Elliptical galaxies in groups and clusters are
predicted to be significantly older than isolated ellipticals. In the
standard CDM model, bright ellipticals in clusters and groups have
$V$-luminosity weighted mean stellar ages that are 3--4 Gyr older than
isolated ellipticals (Kauffmann 1996). The fact that galaxies form
earlier in high density environments is a direct consequence of the
statistical properties of the peaks of Gaussian random fields (Bardeen et
al 1986) and is a prediction of {\em all} models based on the standard
theory.

Another quantity of interest in the number of progenitor galaxies that
merge together to form an elliptical. Kauffmann \& Charlot (1998a) show
that there is a weak trend in the number of progenitors as a function of
the mass of the elliptical: E galaxies with mass $10^{10} M_{\odot}$
typically form from 2--3 progenitors, whereas E galaxies more massive
than $10^{11} M_{\odot}$ form from 5--7 progenitors. Because the number
of progenitors only increases by a factor of $\sim 2.5$ over more than an
order of magnitude in mass, these results also imply that massive Es also
form from the merging of {\em more massive progenitors}. As discussed by
Kauffmann \& Charlot (1998a), this provides an explanation for why a
correlation between mass and metallicity is preserved in a scenario where
ellipticals form via the mergers of disk galaxies. In the models, massive
disks are metal rich because the energy released from supernova
explosions is not sufficient to eject enriched out of the galaxy.
In low-mass disks, most of the metals
escape into the surrounding halo and the galaxy is metal-poor. The
massive, metal-rich disks then merge to form the most massive
ellipticals.

\section {Confrontation with Observations}
The most direct test of the merger picture is the prediction that the
global abundance of bright elliptical galaxies decreases at high
redshift. In all currently popular theoretical models, the abundance of
ellipticals decreases by at least a factor 3 by $z=2$. Because
ellipticals dominate the bright end of the $K$-band luminosity function,
Kauffmann \& Charlot (1998b) have suggested that $K$-band selected redshift
surveys covering a wide area of the sky to a limiting magnitude of 
$K\sim\-20$ would a very effective means of testing the merger scenario.

Testing the merger picture using ellipticals in clusters at high redshift
is much more tricky because of the inherent biases predicted by the
theory. In particular, the dense central regions of the most massive
clusters always contain the oldest galaxies at any redshift. Kauffmann \&
Charlot (1998a) demonstrate that the apparent ``passive evolution''
observed for ellipticals in clusters out to $z \sim 1$ (Stanford,
Eisenhardt \& Dicksinson 1998) is quite consistent with the predictions
of a high-density CDM model. The agreement comes about not because
ellipticals are passively evolving, but because  the observational
selection procedure favours the oldest systems.

It is also possible to test the merger picture by looking for trends in
the ages of elliptical galaxies as a function of environment. This test
has been done at low redshift using a range of different stellar age
indicators. Rose et al (1994),
using a variety of different stellar absorption indices, and Mobasher \&
James (1996), using measurements of the CO absorption feature at 2.3mm,
find evidence that ellipticals in low-density environments contain a
substantial intermediate age population. Bernardi et al (1998), using
the Mg$_2$ index,  find a smaller difference between cluster
and field ellipticals.


\section{Element abundance ratios}
While colours and Balmer line strengths of galaxies are used to infer ages
of  stellar populations, $\alpha$/Fe element ratios
carry information about the timescale over which star formation took
place. The reason for this lies in the delayed Fe enrichment from
Type Ia supernovae, the progenitors of which are believed to be low-mass
and hence long-lived binary stars. Models that incorporate this delay 
succeed in reproducing the
abundance patterns in the solar neighbourhood (e.g.\ Greggio \& Renzini
1983; Matteucci \& Greggio 1986). Although we cannot measure abundance
ratios in elliptical galaxies directly, the analysis of Mg and Fe line
indices in ellipticals (Burstein et al 1984) using population
synthesis models indicates that the ratio Mg/Fe in these objects is
significantly super-solar by 0.2--0.3 dex (Peletier 1989; Worthey, Faber
\& Gonz\'{a}lez 1992). Subsequent studies confirm that this so-called
$\alpha$-enhancement in the stellar populations of ellipticals extends 
out to the effective radius (Davies, Sadler \& Peletier 1993; Carollo \&
Danziger 1994; Fisher, Franx \& Illingworth 1995; Mehlert et al 1999).
In a more quantitative study using $\alpha$-enhanced stellar tracks,
Weiss, Peletier \& Matteucci (1995) derive
$0.3\leq\-[\alpha/{\rm\-Fe}]\leq\-0.7$~dex and in the bulge of our
Galaxy, McWilliam \& Rich (1994) find metal-rich stars having super-solar
O/Fe and Mg/Fe ratios. Depending on the slope of the assumed initial mass
function (IMF), such values require star formation timescales around
$10^8$--$10^9$ years (e.g.\ Matteucci 1994; Thomas, Greggio \& Bender
1999).

In models of hierarchical galaxy formation, star formation in ellipticals
typically does not truncate after 1 Gyr, but continuous to lower redshift
(Kauffmann 1996). It is therefore questionable if hierarchical clustering
would lead to significantly $\alpha$-enhanced giant ellipticals (Bender
1997). Thomas et al (1999) discuss the abundance ratios associated with
mergers of evolved galaxies, and show that a late merger is incapable in
producing significantly super-solar Mg/Fe ratios in the stellar
population. This result agrees qualitatively with a study by Sansom \&
Proctor (1998), who show that Mg/Fe ratios are lower when the object
forms by subsequent mergers.

So far, semi-analytic models have not considered this constraint. In a
preliminary study, Thomas (1999) demonstrated that the {\em average} star
formation history (hereafter SFH) of cluster ellipticals predicted by the
models (see Fig.~\ref{fig:sfh}) led to ${\rm [Mg/Fe]}\sim 0.04$ dex, a
value well below the observational estimate. In the present paper, we use
the {\em individual} SFHs given by the models to explore the distribution
of Mg/Fe values in both cluster and field galaxies of different
morphological types and luminosities.

The prescription of the chemical evolution is explained in Thomas (1999,
and references therein). The chemical evolution code, particularly the
supernova rates of both types, have been calibrated on the abundance
patterns in the solar neighbourhood (Thomas, Greggio \& Bender 1998).
The semi-analytic SFHs are input directly into this code and include both
quiescent star formation in the disk galaxy progenitors as well as the
starburst triggered by the merger. Note that the chemical evolution models
assume that galaxies evolve as single-body, ``closed-box'' systems and
this is not the case in the hierarchical models, where supernova feedback
effects are thought to be important in driving chemical evolution.
Feedback effects could either drive the derived $\alpha$/Fe ratios up or
down, depending on the relative efficiencies with which the
products of Type II and Type Ia supernovae are
expelled. We plan to explore this in detail in future work. We also
assume a universal Salpeter IMF slope $x=1.35$. In the following,
${\rm\-[Mg/Fe]}\equiv\log({\rm\-Mg/Fe})-\log({\rm\-Mg/Fe})_{\odot}$
denotes the global element ratio in the {\em stellar population} of a
individual galaxy. It should be emphasized that our plots do not show
Mg/Fe ratios for the stars inside individual galaxies, but the variation
obtained from one galaxy to another. We also only consider element ratios
at redshift zero. The evolution of Mg/Fe with redshift is discussed in
Thomas (1999). We show results for a cold dark matter power spectrum with
$\Omega=1$, $H_0=50$ km s$^{-1}$ Mpc$^{-1}$, and $\sigma_8=0.67$.

\subsection{Environmental effects}
\label{environment}
\begin{figure}[ht!]
\plotone{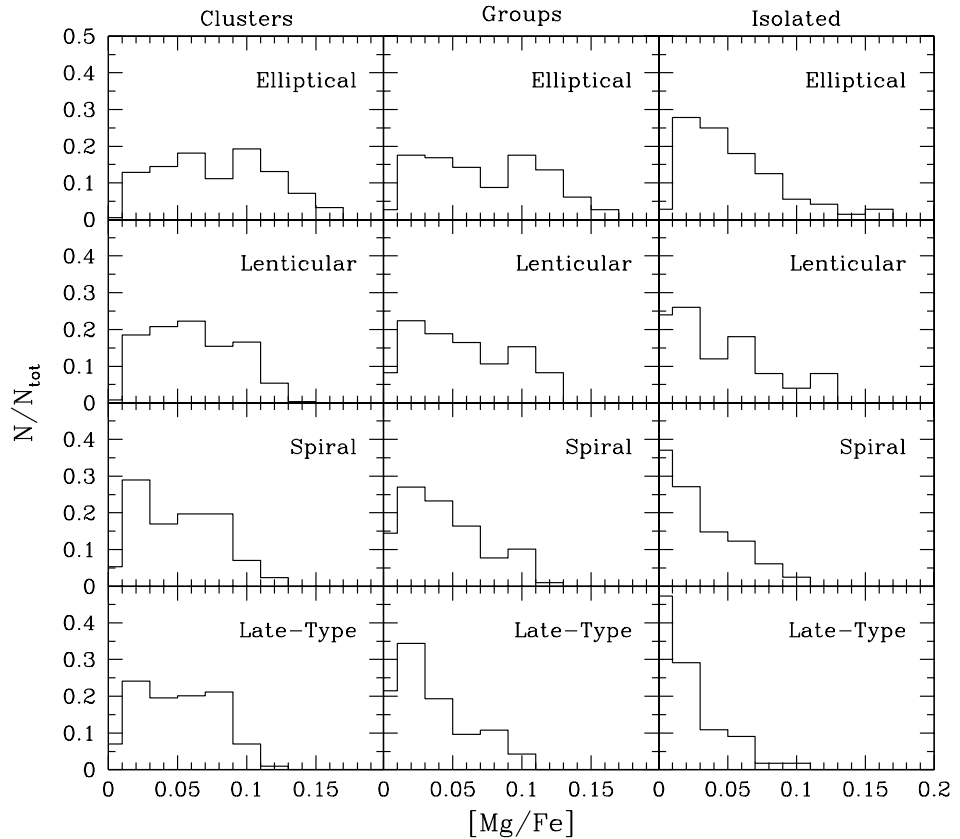}
\caption{Number distribution of global Mg/Fe ratios among galaxies of
different morphologies in environments of varying density.}
\label{fig:distrib}
\end{figure}

Fig.~\ref{fig:distrib} shows the number distribution of Mg/Fe ratios
among galaxies of different morphologies in environments of varying
density. These distributions are computed by averaging over the galaxies
in 15 halos with $V_c= 1000$ km s$^{-1}$ (clusters), 50 halos with
$V_c=500$ km s$^{-1}$ (groups), and 100 halos with $V_c=300$ km s$^{-1}$ (small
groups).

The Mg/Fe values generally scatter between $0\leq {\rm [Mg/Fe]}\leq 0.16$
dex, with a mean of $0.05$--$0.08$ dex, significantly below the observed
estimate quoted above. The spread in Mg/Fe results from the scatter in
star formation timescales in the models (Kauffmann 1996). Late type
galaxies experience more extended SFHs. As a result, these objects have
younger stellar populations and exhibit lower Mg/Fe ratios, in
qualitative agreement with observations. Although elliptical galaxies
have the highest values of Mg/Fe, they still do not match the observed
$\alpha$-enhancement. Only oldest galaxies, which form most of their
stars in the first 1--2 Gyr, reach Mg/Fe close to the lower limit set by
observations. Lenticular galaxies are expected to be slightly less
$\alpha$-enhanced than ellipticals.

Galaxies in small halos form their stars over a more extended timescale
than galaxies in clusters, where the supply of infalling gas is cut off
as soon as the galaxy is accreted. As a result, lower Mg/Fe values are
found for galaxies in small groups. In clusters, the formation
is boosted towards higher redshift and more galaxies form on short
timescales. Bright, isolated ellipticals are thus predicted to have
lower Mg/Fe values than their counterparts in clusters (see also Thomas
1999). Kauffmann \& Charlot (1998a) find that the $V$-light weighted age
difference between bright cluster and ``field'' ellipticals is $\sim 1$
Gyr. This is in good agreement with the estimate by Bernardi et al
(1998), who find that variations in zero-point and slope of the
Mg-$\sigma$ relation of field and cluster ellipticals translate into an
age difference of $1.2\pm 0.35$ Gyr. Their study lacked information from
Fe line indices and they could not determine whether there were trends in
$\alpha$-enhancement from cluster to field. More work in this area would
be very valuable.

\subsection{Trends with luminosity}
\label{luminosity}
\begin{figure}[ht!]
\plotone{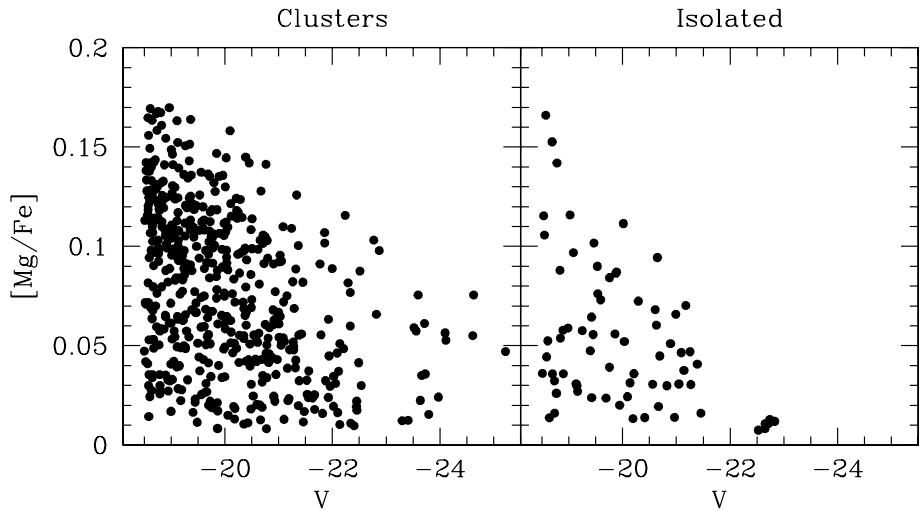}
\caption{Global Mg/Fe in the model ellipticals as a function of absolute 
$V$-magnitude in a high density (left-hand panel) and low-density
(right-hand panel) environment.}
\label{fig:lum}
\end{figure}

In Fig.~\ref{fig:lum} we plot the Mg/Fe values of model ellipticals as a
function of $V$-magnitude for cluster ellipticals (left-hand panel) and
isolated ellipticals (right-hand panel). 

Faint ellipticals ($M_V>-20$) exhibit a scatter in Mg/Fe over a range
from $0-0.17$ dex.  Both the scatter and the median value decrease for
brighter ellipticals because these objects form later and have younger
mean stellar ages. According to Kauffmann \& Charlot (1998a), even though
bright ellipticals are younger than faint ones, they still have redder
colours because they are more metal-rich. Worthey et al (1992) claim,
however, that the Fe index remains constant for elliptical galaxies of
different mass. The tight correlation between Mg index and velocity
dispersion $\sigma$ (Bender, Burstein \& Faber 1993) then implies that
Mg/Fe is higher in more massive ellipticals. This is in clear
disagreement with the trend shown in Fig~\ref{fig:lum}. Note that if the
Fe index is a reasonable tracer of metallicity, the results of Worthey et
al imply that both the Mg-$\sigma$ relation and the colour-magnitude
relation may simply be an effect of an increase in $\alpha$-element
enhancement as a function of the mass of the elliptical. As we have
shown, this is not easily accommodated in hierarchical models, because it
is the {\em low mass} ellipticals that form on short timescales. Another
interpretation is that Fe indices are not viable metallicity indicators
(Gonz\'{a}lez 1993), and the super-solar $\alpha$/Fe ratios in luminous
ellipticals actually result from an Fe deficiency (Buzzoni, Mantegazza \&
Gariboldi 1994). It may be possible to achieve this in models where metal
ejection is treated in detail.
\begin{figure}[ht!]
\plotone{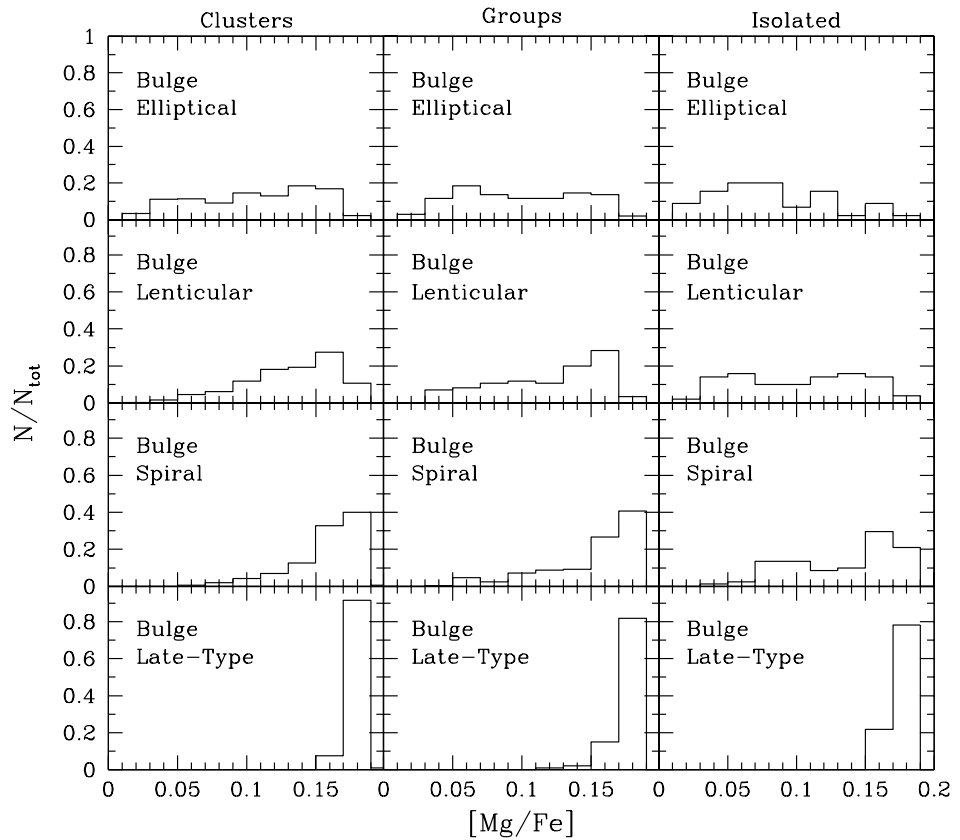}
\caption{Number distribution of Mg/Fe ratios among bulges in
galaxies of different morphologies in environments of varying density.}
\label{fig:bulges}
\end{figure}

\subsection{Bulges}
\label{bulges}
Bulges are the first spheroids to form in major mergers at high redshift
(Kauffmann 1996). Depending on the amount of gas accreted subsequently,
these spheroids may become the bulges of spiral galaxies. As discussed
previously, the classification into morphological type is made according
to disk-to-bulge ratios, so even ellipticals have a minor disk
component.

Fig.~\ref{fig:bulges} shows the distribution of Mg/Fe among the {\em
bulge components} of galaxies of different types in a variety of
environments. The drop of the Mg/Fe ratio from high to low density
environment is less pronounced in the bulge component. Another striking
result is that the bulges of late-type galaxies are more
$\alpha$-enhanced and exhibit a much smaller scatter in Mg/Fe. This
result is expected in a scenario where the bulge forms first and the disk
slowly accretes over a Hubble time. If the bulge is formed from the disk,
one would not expect to see such an effect. It would be interesting to
study Mg and Fe line indices in the bulges of galaxies of different
morphological type in order test some of these trends (see Trager,
Dalcanton \& Weiner 1999).

\begin{figure}[ht!]
\begin{center}
\begin{minipage}{0.8\textwidth}
\plotone{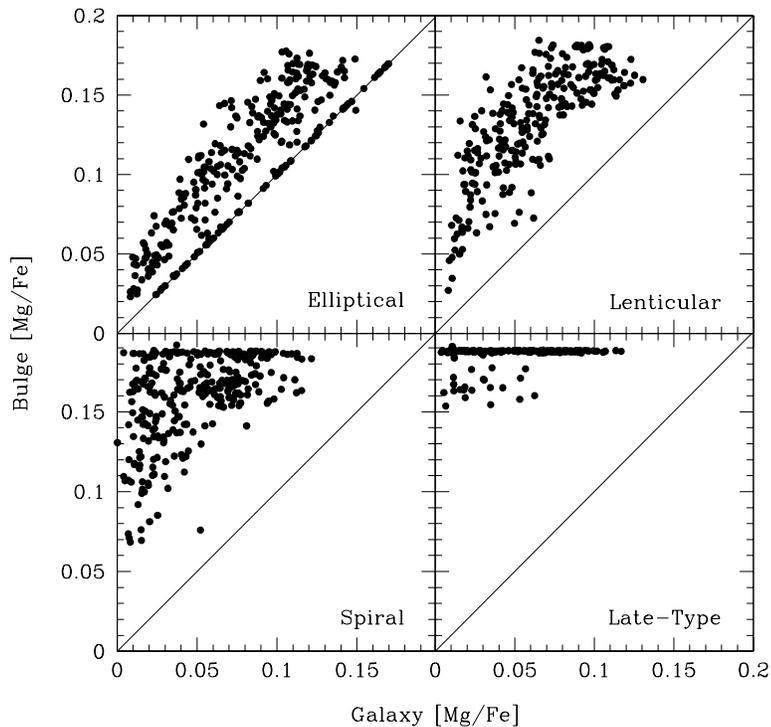}
\end{minipage}
\end{center}
\caption{Mg/Fe ratios in bulges of different galaxy types as a function
of the global Mg/Fe in the respective galaxies. Rich cluster environment.}
\label{fig:bg}
\end{figure}

In Fig.~\ref{fig:bg} we plot the global value of Mg/Fe versus the Mg/Fe
of the bulge component for different types of galaxies in a rich cluster.
The value of Mg/Fe is always higher in the bulge than it is on average
and this difference increases for late type galaxies. This result is
interesting because the bulge component is expected to be more spatially
concentrated, which implies a radial gradient in Mg/Fe inside the galaxy.
In elliptical galaxies, observations indicate Mg/Fe to be constant with
radius inside the half-light radius (Worthey et al 1992; Davies et al
1993; Fisher et al 1995).

\subsection{Caveats and future improvements to the models}

\mbox{}

1) Although we have adopted the SFHs predicted by the semi-analytic models,
we have not yet treated chemical evolution in consistent way.            
Inclusion of metal ejection will allow us to study abundance ratios in
the intra-cluster medium (e.g. Mushotzky et al 1996)

2) Indices are measured close to the centres of the galaxies where      
the starburst population may dominate the light (Barnes \& Hernquist 1996). The
quantities calculated in this work are for the global stellar population.
We also plan to explore two-component models where chemical evolution in
the burst is treated separately.

3) We assumed a universal Salpeter IMF. 
It would be  interesting to study the effect of flattening the IMF of the
stars that form in the burst, which
would increase production of $\alpha$-elements (see Thomas 1999).

4) A more quantitative comparison between model predictions and
observables can be obtained by transforming the calculated Mg/Fe element
ratios to line indices on the basis of Mg/Fe-dependent fitting functions
and stellar tracks.

5) The choice of cosmology has an impact on the SFHs and therefore also
on the resulting abundance ratios. In particular, in low-$\Omega$
universes star formation is pushed towards higher redshift yielding
higher Mg/Fe.


\end{document}